\documentclass[11pt,a4paper]{article}
\usepackage{jcappub}

\newcommand {\be}{\begin{equation}}
\newcommand {\ee}{\end{equation}}
\newcommand {\ba}{\begin{eqnarray}}
\newcommand {\ea}{\end{eqnarray}}

\title{A novel method to extract dark matter parameters from neutrino telescope data}
\author[a,b]{Arman Esmaili} \author[c]{Yasaman Farzan}
\affiliation[a]{Instituto de Fisica Gleb Wataghin - UNICAMP, 13083-859, Campinas, SP, Brazil} \affiliation[b]{School of Particles and Accelerators, Institute for Research in Fundamental Sciences (IPM), P.O.Box 19395-5531, Tehran, IRAN} \affiliation[c]{School of Physics, Institute for Research in Fundamental Sciences (IPM), P.O.Box 19395-5531, Tehran, IRAN}
\emailAdd{arman@ipm.ir} \emailAdd{yasaman@theory.ipm.ac.ir}
\abstract{Recently it has been shown that when the Dark
Matter (DM) particles captured in the Sun directly annihilate into
neutrino pairs, the oscillatory terms in the oscillation
probability do not average to zero and can lead to a seasonal
variation as the distance between the Sun and Earth changes in
time. In this paper, we explore this feature as a novel method to
extract information on the properties of dark matter. We show that
by studying the variation of the flux over a few months, it would
in principle be possible to derive the DM mass as well as new
information on the flavor structure of the DM annihilation modes.
In addition to analytic analysis, we present the results of our
numerical calculations that take into account scattering and
regeneration of neutrinos traversing the Sun.}

\begin{document}
\maketitle

\section{Introduction\label{intro}}

If the dark matter is composed of Weakly Interacting Massive
Particles (WIMPs), it can be trapped inside the Sun and give rise
to a neutrino flux which is in principle detectable at the
neutrino telescopes. Detection of such a flux is an indirect way
of detecting DM which has received much attention in the recent
years. If the flux is high enough, in addition to establishing
WIMP as the DM, information on DM annihilation can be derived from
the properties of the flux such as energy spectrum or flavor
composition.

Recently it has been shown in \cite{Arman} that if the
annihilation to neutrino pairs is one of the dominant modes, the
oscillatory terms in the oscillation probability will lead to a
seasonal variation of the number of  muons produced by the Charged
Current (CC) interaction of $\nu_\mu$ from the Sun in the neutrino
telescopes. That is because in this case, the spectrum of
neutrinos is monochromatic so, contrary to the general belief, the
oscillatory terms do not average to zero and vary throughout a
year as the distance between the Sun and Earth changes due to
nonzero eccentricity of the Earth's orbit. As pointed out in
\cite{Arman}, the seasonal variation can be considered as an
independent tool to derive information on the DM annihilation
modes. In this paper, we further elaborate on this possibility. We
analyze the information that can be derived from this observable
on the flavor structure of DM annihilation amplitude. We also
present numerical results using a code that solves evolution
equations describing the neutrino propagation, taking into account
the effects of neutrino absorption, scattering and $\nu_\tau$
regeneration inside the sun.

Our particular attention is given to the case that DM particles
primarily annihilate into neutrinos. As has been discussed in
detail in ref.~\cite{Lindner}, wide classes of models can be built
in which the dominant annihilation mode of DM particles is the
annihilation into neutrinos. At first sight it might seem that
within such models, the DM  interactions with nuclei will be too
weak to give rise to a  substantial DM abundance in the Sun center
and thus to a significant neutrino flux. We briefly discuss this
issue and show that, despite dominantly annihilating into
neutrinos, DM can have large enough spin-dependent scattering
cross section off the protons inside the Sun. Our main approach in
this paper is however model independent. Our goal is to find out
to what extent the properties of DM (e.g., Br$({\rm DM} +{\rm
DM}\to \nu_\alpha \nu_\beta$)) can be determined by combining the
information on seasonal variation and on flavor composition (more
precisely,  the ratio of the detected muon-like events to
shower-like events).

This paper is organized as follows. In section~\ref{flux}, the
general properties of the flux is described and the seasonal
variation is quantified. In section~\ref{motive}, the theoretical
framework within which a DM pair dominantly annihilates to a
neutrino pair is discussed and shown that the DM capture rate in
the Sun and the subsequent neutrino flux can be high enough to
make the method presented in this paper viable. In
section~\ref{icecube}, the properties and limitation of neutrino
detectors are described and the observable quantities are
formulated. In section~\ref{num}, the numerical code that has been
developed to carry out the calculation is described. In
section~\ref{complementary}, the information that can be in
principle derived from the observable quantities defined in the
previous sections on the flavor structure of DM interactions are
analyzed. In section~\ref{results}, numerical results are
presented and the observed patterns are analyzed.  Concluding
remarks are given in section~\ref{conclusion}.

\section{Neutrino flux from dark matter annihilation in the Sun\label{flux}}

DM particles propagating in the space between stars and planets in
the galaxy can enter the compact objects such as the Sun or the
Earth. If the scattering cross section of DM off nuclei is large
enough, these particles can lose energy at scattering and fall
into the gravitational potential of the Sun or  the Earth. As a
result, the DM particles will be accumulated in the center. The
accumulated DM particles annihilate with each other and produce
the Standard Model (SM) particles. Among the SM particles only
neutrinos can reach the surface from the Sun or Earth center. The
rest will either be trapped or decay before reaching the surface.
DM annihilation can give rise to a neutrino flux either directly
(i.e., ${\rm DM} +{\rm DM}\to \nu \nu$) or as secondaries
(i.e., ${\rm DM} +{\rm DM}\to X \bar{X}$ and subsequently
$X\to \nu Y$). If the mass of the DM particles is above a few
hundred GeV, the neutrino flux from the DM annihilation can be
detectable at the neutrino telescopes such as IceCube. The number
of the neutrino events at IceCube produced by DM annihilation in
the Sun center can be of the order of a few hundred events per
year. The background neutrinos pointing towards the Sun is below
100 events per year \cite{SolarAtmospheric,Honda,wimpless}, so the
detection of an excess of these neutrinos at these energy ranges
can be interpreted as a conclusive indirect detection of DM.

Depending on the couplings and characteristics of the DM
particles, the flavor composition and the energy spectrum of the
neutrino flux can be different. In particular, if neutrinos are
secondaries, we expect a continuous spectrum. On the contrary, if
DM particles annihilate directly into neutrinos, the neutrino
energy spectrum at production point will be monochromatic. That is
because the DM particles annihilate non-relativistically. In fact,
the thermal motion of DM particles can widen the line to a very
narrow Gaussian with width $\delta E/E \sim 10^{-4}$~\cite{Arman}.
Propagation of the monochromatic neutrinos from the dark matter
annihilation in the Sun has been studied in a number of papers
including in \cite{mono,blennow,Barger}. This difference in
spectrum can in principle be invoked to discriminate between the
scenarios predicting different decay modes. Ref.~\cite{Strumia}
has systematically studied the possibility to extract ${\rm
Br}({\rm DM} +{\rm DM}\to \nu \nu$) and ${\rm Br}({\rm DM} +{\rm
DM}\to \tau \bar{\tau}$) by measuring the energy spectrum.

The spectrum of neutrinos emerging from the Sun surface will
consist of a sharp line, corresponding to un-scattered neutrinos,
superimposed on a continuous tail corresponding to scattered or
regenerated neutrinos. The sharp line can still give rise to
oscillatory behavior which is the base of the method suggested in
this paper to determine if $({\rm DM}+{\rm DM}\to \nu \nu$) is
among the dominant annihilation modes. Following
ref.~\cite{Arman}, we define the seasonal variation, $\Delta$,  as
\be \label{deltaEQUIV} \Delta(t_1,\Delta t_1; t_2,\Delta t_2)
\equiv {\tilde{N}(t_1,\Delta t_1)-\tilde{N}(t_2,\Delta t_2) \over
\tilde{N}(t_1,\Delta t_1)+\tilde{N}(t_2,\Delta t_2)} \ee where \be
\label{Ntilde} \tilde{N}(t,\Delta t)\equiv {\int_t^{t+\Delta t}
(dN_\mu/dt) dt \over \int_t^{t+\Delta t}
A_{eff}(\theta[t])L^{-2}(t) dt}\ee in which $L$ is the distance
between the Sun and the Earth and $A_{eff}(\theta[t])$ is the
effective area of the detector which depends on the angle between
the beam direction and the axis of the detectors, $\theta(t)$. As
the Earth orbits around the Sun, $L$ and $\theta$ both change with
time during a year. Notice that with this definition, the $1/L^2$
dependence of the flux does not affect $\Delta$. $\Delta$ vanishes
when the oscillatory terms are absent or averaged out. As shown in
\cite{Arman}, $\Delta$ can be substantially large and exceed 50~\%
so with a moderate accuracy, its nonzero value, which is a proof
of the existence of the mode $({\rm DM} +{\rm DM}\to \nu \nu$) can
be established. The value of $\Delta$ depends on the DM mass,
${\rm Br}({\rm DM} +{\rm DM}\to \nu \nu$) and the initial flavor
composition as well as neutrino oscillation parameters. As pointed
out in ref.~\cite{Arman}, the value $\Delta$ can be used to
extract information on the initial flavor composition, especially
when it is combined with energy spectrum measurements. {Of course,
to construct $\Delta$, the chosen intervals, $\Delta t_1$ and
$\Delta t_2$, should not be too small. Otherwise, the statistics
will be too low, making the derivation of $\Delta$ meaningless due
to the large statistical error. For the energy interval of our
interest ({\it i.e.,} $E_\nu \sim 100-500$~GeV) even  if we take
$\Delta t_1+\Delta t_2=1$ year, the oscillatory terms do not
average out. Even the oscillatory terms corresponding to $\Delta
m_{31}^2$ survive.
We will elaborate on this point  later on. Considering this fact,
there is practically no upper bound on $\Delta t_i$, except that
$\Delta t_1+\Delta t_2<1$ year.}

\section{Theoretical motivation \label{motive}}

As discussed in ref.~\cite{Arman}, in order to have non-vanishing
$\Delta$, the following two conditions are necessary: (1) The
neutrino spectrum has to include a sharp line with width $\delta
E/E\ll 4\pi E/(\Delta m^2L)$ which can be realized if the DM annihilation into neutrino pair is significant; (2) the neutrino flux at production must be non-democratic;  i.e.,
$F_{\nu_e}^0: F_{\nu_\mu}^0: F_{\nu_\tau}^0 \ne 1:1:1$.
We have already discussed the reason for the first condition. The
second condition is also necessary because for $F^0\equiv
F_{\nu_e}^0= F_{\nu_\mu}^0= F_{\nu_\tau}^0$, the neutrino flux at
Earth will be independent of the oscillation probability ($
F_{\nu_\beta}=\sum_\alpha F^0_{\nu_\alpha}P(\nu_\alpha \to
\nu_\beta)=F^0\sum_\alpha P(\nu_\alpha \to \nu_\beta)= F^0 $) and
as a result, $\Delta$ will vanish  because  it comes from the
oscillatory terms in the oscillation probability.

As discussed in ref.~\cite{Lindner}, various classes of models can
be built that satisfy both these conditions. An explicit example
is given in ref.~\cite{Farzan}. In order to measure $\Delta$, the
number of collected neutrino events has to be larger than a few
hundred. If the flux is close to the upper bound from AMANDA, such
an amount of data can be collected at IceCube. For $m_{DM} \sim
100 ~{\rm GeV}$, the neutrino emission from the Sun with a rate of
order of $10^{20}~{\rm sec}^{-1}$ will be enough to lead to a few
hundred events per year. In the saturation limit where the
capture rate of DM particles by the Sun, $C_\odot$, equals the
neutrino emission rate (i.e., two times the annihilation rate),
the neutrino emission rate can be translated into a bound on the
DM-nuclei cross section. As seen from eqs.~(9.25) and (9.26) of
\cite{review}, for spin-independent scattering mediated by a
scalar, with a DM-nucleon cross section, $\sigma_n$, as small as
$10^{-9}$~pb, $C_\odot=10^{20}~{\rm sec}^{-1}$ can be achieved.
Such cross section is below the present bound on spin-independent
DM-nucleon cross section \cite{SI}. It is noteworthy that while the Sun
is mainly composed of protons, for spin-independent interactions,
the heavier nuclei inside the Sun play the dominant role in
capturing the DM particles. That is because the nucleons in a
nucleus interact coherently with DM so the cross section grows
quadratically with nucleus mass. If the DM-quark interaction is
mediated via a heavy scalar, we would have \be {\langle
\sigma({\rm DM}+{\rm DM}\to q \bar{q}) v\rangle|_{{\rm
decoupling}} \over \sigma_n}\sim \left({m_p+m_{DM}\over
m_p}\right)^2\sim 10^{4}.\ee Thus, $\langle \sigma({\rm DM}+{\rm
DM}\to q \bar{q}) v\rangle|_{{\rm decoupling}}\sim 10^{-5}~pb \ll
\langle \sigma_{tot} v \rangle \sim 1~pb$. The above discussion
shows that it is possible to construct a model within which DM-nucleus interaction is large enough for sufficient DM capture in the Sun but at the same time, the main annihilation mode of DM
pairs is the annihilation into neutrino pairs.

\section{The IceCube neutrino telescope\label{icecube}}

\subsection{General description}

IceCube is a km$^3$-scale neutrino telescope recently
completed at the south pole \cite{icecube}. IceCube, which
encompasses the AMANDA experiment \cite{amanda}, has been designed
to detect the flux of galactic and extragalactic neutrinos with
energies higher than a few 10 GeV. The detector consists of 80
strings, each carrying 60 photo-multiplier tubes (PMT) which are
being deployed between 1450 m and 2450 m depth. In addition, there
are six more strings with a denser array of PMTs between 1760 m
and 2450 m, which with the surrounding strings constitute the
DeepCore detector at the far depth of the IceCube. Since the PMTs
at the DeepCore are closer, the sensitivity of DeepCore is higher
than the rest of IceCube and its detection threshold is lower.
As explained later, the background events at the DeepCore
are also highly suppressed such that, in spite of the small volume
of DeepCore which  reduces the statistics of the signal, the
discovery chance can still be comparable to that by the whole
IceCube.

For neutrinos with energy ${\mathcal{O}} (100\, {\mathrm{GeV}}) $,
IceCube cannot distinguish between the neutrino flavors
completely. Practically, IceCube can identify two types of events:
$\mu$-track and shower-like events. Let us explain them one by
one. The $\mu$-track events are the Cherenkov radiation collected
by the PMTs when a muon passes through the instrumented volume of
the detector. The muons can be produced in two ways: i) from the
CC interaction of $\nu_\mu$ and $\bar{\nu}_\mu$; ii) from CC
interactions of $\nu_\tau$ (and $\bar{\nu}_\tau$) which produces
the tau (and the anti-tau) particle and the subsequent $\tau^- \to
\mu^- \bar{\nu}_\mu \nu_\tau$ (and $\tau^+ \to \mu^+ {\nu}_\mu
\bar{\nu}_\tau$). However, the small branching ratio of the tau
particle's leptonic decay, Br$(\tau\to \mu\nu\nu)=17 \%$, makes
the contribution from $\nu_\tau$ and $\bar{\nu}_\tau$ subdominant.
Thus, the number of $\mu$-track events is practically given by the
fluxes of $\nu_\mu$ and $\bar{\nu}_\mu$ incident on the IceCube.
Considering the position of the muon production point, the
$\mu$-track events for the whole IceCube can be divided into two
categories: i) Through-going $\mu$-track events for which the
muons are produced inside or in the vicinity of the instrumented
volume of the detector; ii) Contained $\mu$-track events for which
the production point is within the geometrical volume of the
detector.
Similar consideration also holds for the muon neutrinos detected
at the DeepCore. In fact, tracks originated inside the IceCube
(but outside DeepCore) and passing through the DeepCore will count
as ``through-going" events for DeepCore and as ``contained" events
for IceCube.

The second type of events that can be identified at the IceCube is
the shower-like events which are nearly spherical volumes of
Cherenkov radiation from the hadronic or electromagnetic cascades.
The interactions contributing to the shower-like events consist
of: i) Neutral Current (NC) interaction of all the three neutrinos
flavors $\nu_\alpha$ and $\bar{\nu}_\alpha$; ii) CC interaction of
$\nu_e$ and $\bar{\nu}_e$; iii) CC interaction of $\nu_\tau$ and
$\bar{\nu}_\tau$, and the accompanying hadronic decay of the
produced tau particle. The event rate calculation of these
interactions can be found in \cite{esmaili}.

\subsection{Observable quantities at neutrino telescopes}

In this subsection, we quantify the observable quantities that we shall
use in our subsequent analysis; namely, the seasonal variation and
the ratio of $\mu$-track events to shower-like events. We  define
two versions of the observable $\Delta(t_1,\Delta t_1; t_2,\Delta
t_2)$ in eq.~(\ref{deltaEQUIV}). \begin{itemize}
\item The $\Delta^{\mathrm{IC}}(t_1,\Delta t_1; t_2,\Delta t_2)$
is the seasonal variation of the through-going $\mu$-track events
for the whole IceCube. To compute $\Delta^{\mathrm{IC}}$, we
calculate the number of $\mu$-track events (per unit of time) with
the following formula
\begin{align}\label{through}
\frac{dN^{\mathrm{IC}}_\mu}{dt} &=\int^{m_{DM}}_{E_{\mathrm{thr}}}
\int^{E_{\nu_\mu}}_{E_{\mathrm{thr}}}\frac{d\Phi_{\nu_\mu}}{dE_{\nu_\mu}}
\left[ \frac{d\sigma^{CC}_{\nu p}}{dE_\mu}(E_{\nu_\mu}) \rho_p +
\frac{d\sigma^{CC}_{\nu n}}{dE_\mu}(E_{\nu_\mu}) \rho_n \right]
\nonumber\\ &  \nonumber\\ & \times A_{eff}(E_\mu,\theta[t])
\left(R_\mu (E_\mu,E_{\mathrm{thr}})+ d\right) dE_\mu dE_{\nu_\mu}
+(\nu_\mu \to \bar{\nu}_\mu) ,
\end{align}
where $d\Phi_{\nu_\mu}/dE_{\nu_\mu}$ is the flux of $\nu_\mu$ at
the IceCube site, $d\sigma^{CC}_{\nu p}/dE_\mu$ and
$d\sigma^{CC}_{\nu n}/dE_\mu$ are the scattering CC partial cross
section of muon neutrinos off proton and neutron respectively,
$\rho_p\sim 5/9 N_A$ cm$^{-3}$ and $\rho_n\sim 4/9 N_A$ cm$^{-3}$
are the number densities of protons and neutrons in the vicinity
of the IceCube in terms of Avogadro's number $N_A$, $R_\mu$ is the
muon range in the ice \cite{muon-range}, and $d=1$ km is the size
of IceCube.  $A_{eff}$, which is given in appendix of
\cite{eff-area}, is the IceCube's effective area of muon detection
and depends on the zenith angle of the arriving neutrinos
$\theta[t]$. Finally, The $E_{\mathrm{thr}}$ is the energy
threshold of muon detection at the IceCube which we have set equal
to 40 GeV \cite{eff-area}. Notice that in writing
eq.~(\ref{through}), we have taken into account the fact that the
maximum $E_\mu$ equals $E_{\nu_\mu}$.

\item The second version of the observable defined in
eq.~(\ref{deltaEQUIV}) is $\Delta^{\mathrm{DC}}(t_1,\Delta t_1;
t_2,\Delta t_2)$ corresponding to the seasonal variation of the
DeepCore $\mu$-track events. The number of $\mu$-track events
determining $\Delta^{\mathrm{DC}}$ is given by

\begin{align}\label{DCmu}\frac{dN^{\mathrm{DC}}_\mu}{dt} &=
\int^{m_{DM}}_{E_{\mathrm{thr}}}
\int^{E_{\nu_\mu}}_{E_{\mathrm{thr}}}\frac{d\Phi_{\nu_\mu}}{dE_{\nu_\mu}}
\left[ \frac{d\sigma^{CC}_{\nu p}}{dE_\mu}(E_{\nu_\mu}) \rho_p +
\frac{d\sigma^{CC}_{\nu n}}{dE_\mu}(E_{\nu_\mu}) \rho_n \right]
\nonumber \\ & \nonumber \\ & \qquad \qquad\qquad\quad\quad\times
V^{\mathrm{DC}}_{eff}(E_\mu)\, dE_\mu \, dE_{\nu_\mu} +(\nu_\mu
\to \bar{\nu}_\mu) , \end{align} where $V^{\mathrm{DC}}_{eff}$ is
the effective volume of the DeepCore given by \cite{wimpless}
\begin{equation}
V^{\mathrm{DC}}_{eff}(E_\mu)=\left( 0.0056 \log E_\mu+0.0146
\right) \Theta(275-E_\mu)+0.0283\, \Theta(E_\mu-275),
\end{equation} in which $\Theta(x)$ is the Heaviside step function
and $E_\mu$ is in GeV unit. We have set also the energy threshold
of the muon detection at the DeepCore equal to 40 GeV \cite{dc}.
\end{itemize}

In order to employ the shower-like events in our analysis, we
define the following observable
\begin{equation}\label{Rdefinition}
R\equiv\frac{{\text {Number of $\mu$-track events}}}{{\text
{Number of shower-like events}}}.\end{equation} \\ Similar to the
observable $\Delta$, we define two quantities $R^{\mathrm{IC}}$
and $R^{\mathrm{DC}}$ corresponding to the through-going and
DeepCore events respectively. In the case of $R^{\mathrm{IC}}$,
the numerator of eq.~(\ref{Rdefinition}) will be calculated using
 eq.~(\ref{through}) and for the denominator we use the
geometrical volume of the whole IceCube detector. Similarly, for
the calculation of the numerator of $R^{\mathrm{DC}}$ we use
eq.~(\ref{DCmu}); while in the denominator we use the geometrical
volume of the DeepCore.
Generally, if $E_{\mathrm{thr}}^{\mathrm{shower}}\geq m_{DM}$,
where $E_{\mathrm{thr}}^{\mathrm{shower}}$ is the energy threshold for
the detection of shower-like events, there will be no
shower-like events at the detector.
To carry out  this analysis, we set
$E_{\mathrm{thr}}^{\mathrm{shower}}$ equal to 100~GeV. This
assumption is rather optimistic as in  reality
$E_{\mathrm{thr}}^{\mathrm{shower}}\sim 1$~TeV. However, in this
paper our main focus is on $\Delta$ which does not employ
shower-like events. If no practical way to improve
$E_{\mathrm{thr}}^{\mathrm{shower}}$ is found, there will be even
more motivation for pursuing the measurement of $\Delta$.

\subsection{Background}

In the measurement of both  observables $\Delta$ and $R$ at the
IceCube, the background events should be considered. The two main
sources of backgrounds at the IceCube are atmospheric muons and
atmospheric neutrinos. The atmospheric muons are isotropically
produced in the collision of the cosmic rays with the atmosphere.
However, from March equinox to September equinox (corresponding to
spring and summer in the northern hemisphere) when the Sun is
below the horizon, the Earth acts as a filter for the atmospheric
muons so these backgrounds will be suppressed. Thus, for the
through-going events at the IceCube we integrate
eq.~(\ref{through}) over the March to September period of the
year. For the DeepCore, the instrumented volume of the IceCube
surrounding it acts as a veto (up to one part in $10^6$
\cite{dc,angular}) so it is possible to take data at DeepCore
throughout the whole year. Thus, we calculate the number of
muon-track events at the DeepCore by integrating eq.~(\ref{DCmu})
over the whole year. The other source of backgrounds, atmospheric
neutrinos, can be suppressed thanks to the high angular resolution
of the IceCube in the $\mu$-track reconstruction (about $1^\circ$
\cite{angular}). Considering only the events in a cone with half
angle $1^\circ$ around the position of the Sun will reduce the
number of through-going atmospheric neutrino backgrounds to $\sim
6$ events per year ( $\sim 3$ yr$^{-1}$ for DeepCore). There is
also an irreducible background from the solar atmospheric
neutrinos but the flux is expected to be low
\cite{SolarAtmospheric}.

\section{Numerical calculations\label{num}}

As mentioned in section~\ref{intro}, the neutrinos at the center
of Sun are monochromatic with energy equal to the DM particle
mass. However, due to the NC and CC interaction of the neutrinos
with the Sun medium,  the spectrum of the neutrinos emerging from
the surface of the Sun, in addition to the sharp line of the
monochromatic neutrinos, will contain a continuous tail
corresponding to the scattered or regenerated neutrinos. We have
written a Mathematica code to numerically compute the flux of
neutrinos at the surface of the Sun. In the code we have taken
into account the oscillation of the neutrino flavors in the Sun
medium, CC and NC interactions of the neutrinos with the nuclei,
and the neutrinos coming from the tau regeneration effect.

Flavor oscillation of the neutrinos in the Sun medium has been
computed using the updated matter density profile of the Sun
$n_e(r)$ from ref.~\cite{electrondensity}. The oscillation
Hamiltonian is
\begin{equation}
{\mathcal{H}}_{\mathrm{osc}}=\frac{(M^2)_{{\mathrm{diag}}}}{2E_\nu}
+{\mathrm{diag}}(\sqrt{2}G_Fn_e,0,0),\end{equation} where $E_\nu$
is the neutrino energy, $G_F$ is the Fermi constant and

$$(M^2)_{{\mathrm{diag}}}=U^{\dagger} M^2_\nu U={\mathrm{diag}} (-\Delta m^2_{12},0,\xi
\Delta m^2_{32}).$$ $M_\nu$ is the neutrino mass matrix and $U$ is
the PMNS unitary mixing matrix. The parameter $\xi$ takes values
+1 and -1, respectively corresponding to the normal and inverted
neutrino mass schemes. For the mixing parameters we insert $\Delta
m^2_{12}=8.0\times 10^{-5}$~eV$^2$, $\Delta m^2_{32}=2.4\times
10^{-3}$~eV$^2$, $\theta_{12}=33.46^\circ$ and
$\theta_{23}=45^\circ$, $\theta_{13}=0$ and $7^\circ$,
$\delta=0,\pi/2$ and $\pi$.

For the NC and CC interaction cross sections we have used the
tabulated values from  ref.~\cite{Gandhi}. The NC interaction of a
neutrino flavor lowers the energy of the neutrino without changing
the flavor. The CC interaction of $\stackrel{(-)}{\nu_e}$ and
$\stackrel{(-)}{\nu_\mu}$ transforms them into the corresponding
charged leptons and therefore reduces the neutrino flux. The tau
and anti-tau particles created by   the CC interactions of
$\nu_\tau$ and $\bar{\nu}_\tau$ promptly decay and reproduce
neutrinos (``tau regeneration''). For example, the decay of the
tau particle through the channel
$\tau^-\to\mu^-\bar{\nu}_\mu\nu_\tau$ reinject $\bar{\nu}_\mu$ and
$\nu_\tau$ with lower energy into the flux of neutrinos. We
consider the following decay channels for the tau particle:
Br$(\tau \to e \nu_e \nu_\tau)=0.18$, Br$(\tau \to \mu \nu_\mu
\nu_\tau)=0.18$, Br$(\tau \to \pi \nu_\tau)=0.12$, Br$(\tau \to
a_1 \nu_\tau)=0.13$, Br$(\tau \to \rho \nu_\tau)=0.26$ and
Br$(\tau \to \nu_\tau X)=0.13$. As can be seen, tau regeneration
couples the evolution equations of different flavors together (it
also couples the neutrino flux evolution to the anti-neutrino
evolution). The details of the tau regeneration calculation as
well as the neutrino spectra in the above decay channels can be
found in \cite{regeneration}. The set of coupled evolution
equations describing the neutrino propagation can be found in
\cite{Barger}.

Let us take the state of the neutrinos at the production point to
be $|\nu_\alpha\rangle$. By solving the evolution equations
numerically, we obtain the state of the neutrinos
$|\nu_\alpha;{\rm surface}\rangle$ at the surface of the Sun.
After leaving the Sun, neutrinos will propagate through the empty
space towards the Earth. The evolution of the state
$|\nu_\alpha;{\rm surface}\rangle$ from the Sun surface to the
Earth is given simply by multiplying each mass eigenstate with
phases, $\exp[i m_i^2L/E_\nu]$. These phases depend on the
distance between the Sun and Earth (i.e., $L$) which is a
function of time. Considering the eccentricity of the Earth's
orbit around the Sun and also the tilt of the Earth's rotation
axis with respect to the ecliptic plane, we have calculated
$\theta(t)$ and $L(t)$ during a year (see eq.~\ref{deltaEQUIV} and
\ref{Ntilde}). After calculating the values of $\theta(t)$, $L(t)$
and the spectrum of neutrinos arriving at Earth during the year,
we have computed the observables $R$ and $\Delta$ according to the
formulation presented in section~\ref{icecube}.

It should be mentioned that the flux of neutrinos arriving in  the
Earth has been calculated previously by Blennow et al.
\cite{blennow} for different annihilation modes of DM and the
mixing parameters $\theta_{13}=0,5^\circ,10^\circ$ and $\delta=0$.
However, since we needed the flux of neutrinos for other values of
the mixing parameters (especially $\delta=0, \pi/2$ and $\pi$), we
have calculated it independently. With the same input values of
the mixing parameters, our results are in complete agreement with
the results of \cite{blennow}.

\section{Analysis of information from seasonal variation\label{complementary}}

Let us take the amplitude of ${\rm DM}+{\rm DM} \to \nu_\alpha
+\stackrel{(-)}{{\nu}_\beta}$ to be ${\mathcal{M}}_{\alpha
\beta}$. The state resulting from the DM pair annihilation will be
a ``pure" coherent state of type $|\psi\rangle=\sum_{\alpha \beta}
{\mathcal{M}}_{\alpha \beta}|\nu_\alpha(\vec{p}_1)
\stackrel{(-)}{\nu_\beta}\!\!\!(\vec{p}_2) \rangle$. The density
matrix is $|\psi\rangle \langle \psi
|=\sum_{\alpha\beta\gamma\sigma} \tilde{\rho}_{\alpha \beta,
\gamma\sigma}|\nu_\alpha \stackrel{(-)}{{\nu}}_\beta\rangle\langle
\nu_\gamma \stackrel{(-)}{{\nu}}_\sigma|$ in which
$\tilde{\rho}_{\alpha \beta,\gamma\sigma}={\mathcal{M}}_{\alpha
\beta}{\mathcal{M}}_{\gamma \sigma}^*$. Since the DM pairs are
almost at rest inside the Sun, the momenta of the produced pair
will be opposite to each other (i.e., $\vec{p}_1 \simeq
-\vec{p}_2$) which means at most one of the emitted neutrinos will
reach the earth. As a result, the density matrix of the neutrinos
pointing towards us will be the following ``reduced" matrix
$$ {\rho}_{\alpha \beta}|\nu_\alpha \rangle \langle
\nu_{\beta}|~~~~{\rm with}~~~~{\rho}_{\alpha \beta}=\sum_\gamma
\tilde{\rho}_{\alpha \gamma,\beta
\gamma}=({\mathcal{M}}{\mathcal{M}}^\dagger)_{\alpha \beta}\ .$$
Notice that although the two particle state $|\psi \rangle$ is
pure (i.e., $\tilde{\rho} \log\tilde{\rho}=0$), the neutrino
states reaching the detector are not in general pure (i.e.,
${\rho} \log {\rho}\ne 0$). In case that the annihilation is
lepton number violating ${\rm DM}+{\rm DM} \to \nu_\alpha
+{{\nu}_\beta}$, ${\mathcal{M}}_{\alpha \beta}$ is obviously
symmetric. As a result, ${\rho}$ yields $ {\mathcal{M}}_{\alpha
\beta}$; i.e., for ${\rho}=V \rho_{diag} V^\dagger$ where $V$ is a
unitary matrix, ${\mathcal{M}}=V (\rho_{diag})^{1/2} V^T$. In case
that the DM annihilation is lepton number conserving (i.e., ${\rm
DM}+{\rm DM} \to \nu_\alpha \bar{\nu}_\beta$), up to subdominant
(or perhaps zero) CP-violating effects, we can write
${\mathcal{M}}_{\alpha \beta}={\mathcal{M}}^*_{ \beta\alpha}$.
Under this condition, ${\mathcal{M}}_{\alpha \beta}$ can be again
derived from ${\rho}$: for a Hermitian ${\mathcal{M}}$, ${\rho}=V
\rho_{diag} V^\dagger$ yields
${\mathcal{M}}=V(\rho_{diag})^{1/2}V^\dagger$. From $
{\mathcal{M}}_{\alpha \beta}$, information on flavor structure of
the couplings of DM can be derived. By studying the neutrino flux
there is however no way to figure out whether the DM annihilation
is lepton number violating or conserving. In the following, we
discuss the possibility of constraining ${\rho}$ from the
observable quantities like $R$ and $\Delta$.

The density matrix ${\rho}$ can be diagonalized as \be \label{nNu}
{\rho}=\sum_\alpha n_\alpha |{\nu}_\alpha^\prime \rangle
\langle{\nu}_\alpha^\prime|\ee where $\alpha=1,2,3$, $n_\alpha\geq
0$ and $\langle
{\nu}_\alpha^\prime|{\nu}_\beta^\prime\rangle=\delta_{\alpha
\beta}$. Notice that ${\nu}_\alpha^\prime$ can in general be
different from mass eigenstates $\nu_1$, $\nu_2$ and $\nu_3$ as
well as from flavor eigenstates $\nu_e$, $\nu_\mu$ and $\nu_\tau$.

Once ${\nu}_\alpha^\prime$ arrives at the surface of the Sun, it will be a
different state that can in general be written in terms of the
vacuum mass eigenstates as \be |{\nu}_\alpha^\prime;{\rm
surface}\rangle=a_{\alpha 1} |1\rangle+a_{\alpha 2}
|2\rangle+a_{\alpha 3} |3\rangle \ .\label{timed}\ee We do not a
priori know what is the flavor composition of $\nu_\alpha^\prime$,
so $a_{\alpha i}$ are unknown. A general neutrino state $|\psi
\rangle$ produced at the Sun center, while crossing the Sun,
evolves into $|\psi;{\rm surface} \rangle$ as follows
$$ |\psi;{\rm surface}\rangle
=W(t_P,t_S)|\psi \rangle~~~~{\rm with}~~~~
W(t_P,t_S)=\prod_{t_P}^{t_S} {\mathcal{T}}[1+i H(t) dt]$$ where
$t_P$ and $t_S$ are respectively the production time and the time
that neutrino reaches the surface and ${\mathcal{T}}$ denotes time
ordering. Regardless of the time dependence of $H$, the evolution
matrix is unitary so for any arbitrary states $|\psi\rangle$ and
$|\chi\rangle$  $$\langle \chi;{\rm surface}|\psi;{\rm
surface}\rangle =\langle \chi|\psi\rangle.$$
As a result, the $|{\nu}_\alpha^\prime \rangle$ states after
evolving in time remain perpendicular:
$$\langle {\nu}_\alpha^\prime;{\rm surface}|{\nu}_\beta^\prime;{\rm
surface}\rangle =\sum_i a_{\alpha i}a^*_{\beta i}=\delta_{\alpha
\beta}.$$ Thus, the knowledge of two rows of the $3\times 3$
matrix, $a_{\alpha i}$, is enough to determine the matrix
$a_{\alpha i}$.

On the way to the Earth, the state will evolve into\footnote{The
Earth matter effect is irrelevant because for this energy range,
the effective mixing is negligible \cite{Arman}.} \be
|\nu_\alpha^\prime;{\rm detector}\rangle=a_{\alpha 1}
|1\rangle+a_{\alpha 2} e^{i\Delta_{12}}|2\rangle+a_{\alpha 3}
e^{i\Delta_{13}} |3\rangle \ ,\ee in which $\Delta_{ij}\equiv
\Delta m_{ij}^2 L/(2E)$. Obviously, the number of $\mu$-tracks and
therefore both $\Delta$ and $R$ depend on $\sum_\alpha n_\alpha
P(\nu_\alpha \to \nu_\mu)$. Let us then discuss these oscillation
probabilities \be \label{ppp} P(\nu_\alpha \to \nu_\mu)=\sum_i
|a_{\alpha i}|^2|U_{\mu i}|^2 + \ee
$$ 2\Re [ a_{\alpha 1}^* a_{\alpha 2} U_{\mu 1} U_{\mu 2}^* e^{i
\Delta_{12}}]+2\Re [ a_{\alpha 1}^* a_{\alpha 3} U_{\mu 1} U_{\mu
3}^* e^{i \Delta_{13}}]+2 \Re[a_{\alpha 2}^* a_{\alpha 3} U_{\mu
2} U_{\mu 3}^* e^{i (\Delta_{13}-\Delta_{12})}]\ .$$

As shown in ref. \cite{Arman}, the first oscillatory term given by
$\exp[i\Delta_{12}]$  does not vanish as we integrate on time over
a year when the Earth orbits around the Sun. This term results in
sizeable seasonal variation. Changing the DM mass, which coincides
with the energy of neutrino line, this term rapidly oscillates
with period \be \label{pppperiod}\frac{\delta
m_{DM}}{m_{DM}}=\frac{4 \pi m_{DM}}{\Delta m^2_{21} L}=0.04
\frac{m_{DM}}{200~{\rm GeV}} \ .\ee
It is noteworthy that even the effects of the phase $\Delta_{13}$
do not completely vanish. The reason is that due to angular
momentum conservation, the Earth slows down when it is at the
aphelion so the eccentricity of the orbit gives rise to
non-vanishing average. To quantify this claim, let us define \be
\label{ooo12}{ O}_{12}(t, \Delta t)\equiv{\int_t^{t+\Delta t} e^{i
\Delta_{12}(t)} A_{eff}(t)L^{-2}(t) dt \over\int_t^{t+\Delta t}
A_{eff}(t)L^{-2}(t) dt}, \ee and \be \label{ooo13}{ O}_{13}(t,
\Delta t)\equiv{\int_t^{t+\Delta t} e^{i \Delta_{13}(t)}
A_{eff}(t)L^{-2}(t) dt \over\int_t^{t+\Delta t}
A_{eff}(t)L^{-2}(t) dt} .\ee For $E_\nu \sim 100~$GeV,
$|{O}_{12}|\sim 1$ and $| O_{13}|\sim 0.1$ which means even the
effects driven by $\Delta_{13}$ do not completely average out. The
average oscillation probability can be written as
$$ \langle P(\nu_\alpha \to \nu_\mu)\rangle|_{t}^{t+\Delta
t}\equiv{\int_{t}^{t+ \Delta t}P(\nu_\alpha \to \nu_\mu) A_{eff}
L^{-2}(t) dt \over\int_{t}^{t+ \Delta t} A_{eff} L^{-2}(t)}=$$ \be
\label{average}\sum_i |a_{\alpha i}|^2|U_{\mu i}|^2+2
\Re[a_{\alpha 1}^*a_{\alpha 2} U_{\mu 1}U_{\mu 2}^* O_{12}] +{\cal
O}(a_{\alpha i}^2U_{\mu i}^2O_{13}).\ee

Let us now discuss the antineutrinos. The density matrix of
antineutrinos at production can be written as \be
\bar{\rho}_{\alpha \beta}|\bar{\nu}_\alpha \rangle \langle
\bar{\nu}_{\beta}|~~~~ {\rm with}~~~~~\bar{\rho}_{\alpha
\beta}\equiv \sum_\gamma \tilde{\rho}_{\gamma \alpha,\gamma\beta
}=({\mathcal{M}}^T{\mathcal{M}}^*)_{\alpha \beta}\ . \ee The
density matrix $\bar{\rho}$ can be diagonalized as follows \be
\bar{\rho}=\sum_\alpha n_\alpha |\bar{\nu}_\alpha^\prime \rangle
\langle \bar{\nu}_\alpha^\prime |.\label{nAntinu}\ee Notice that
for the lepton number and CP conserving annihilation (${\rm
DM}+{\rm DM} \to \nu \bar{\nu}$) with
${\mathcal{M}}={\mathcal{M}}^\dagger$, we find
$\rho=\bar{\rho}^*$. On the other hand,  for lepton number
violating annihilation (${\rm DM}+{\rm DM} \to \nu
+\nu,\bar{\nu}+\bar{\nu}$) with ${\mathcal{M}}={\mathcal{M}}^T$,
we find $\rho=\bar{\rho}$. In any case $n_\alpha$ in
eq.~(\ref{nNu}) for neutrinos and the one in eq.~(\ref{nAntinu})
for antineutrinos are the same. Moreover,
$|\nu_\alpha^\prime\rangle$ and $|\bar{\nu}_\alpha^\prime\rangle$
at the source are the charged conjugate of each other. However,
the matter effects on $|\nu_\alpha^\prime\rangle$ and
$|\bar{\nu}_\alpha^\prime\rangle$ are different so
$|\bar{\nu}^\prime_\alpha;{\rm surface}\rangle$ will not in
general be the charged conjugate of $|{\nu}^\prime_\alpha;{\rm
surface}\rangle$.

Let us expand the antineutrino state $\bar{\nu}_\alpha^\prime$ at
the surface as  \be |\bar{\nu}_\alpha^\prime;{\rm
surface}\rangle=\bar{a}_{\alpha 1} |\bar{1}\rangle+\bar{a}_{\alpha
2} |\bar{2}\rangle+\bar{a}_{\alpha 3}  |\bar{3}\rangle
\label{antiExpansion}.\ee As explained above because of the matter
effects on the way from the Sun center to its surface,  $a_{\alpha
i}^* $ is not in general equal to $ \bar{a}_{\alpha i}$.
It is straightforward to confirm that the oscillation
probability for anti-neutrinos, $P(\bar{\nu}_\alpha \to
\bar{\nu}_\mu)$, and its average are given by formulas
respectively similar to Eqs.~(\ref{ppp},\ref{average}) replacing
$a_{\alpha i}$ with $\bar{a}_{\alpha i}$ and $U_{\mu i}$ with
$U_{\mu i}^*$.

Since the scattering cross sections of antineutrinos are smaller than
that of neutrinos by a factor of about 2, their contribution to the
events at the detector will be smaller.

Obviously, replacing $\mu$ with $\tau$, we obtain the formula for
$ \langle P(\nu_\alpha \to \nu_\tau)\rangle|_{t}^{t+\Delta t}$ and
$ \langle P(\bar{\nu}_\alpha \to
\bar{\nu}_\tau)\rangle|_{t}^{t+\Delta t}$. Notice that to the
leading order in $\theta_{13}$ and $\theta_{23}-\pi/4$, $|U_{\mu
i}|\simeq |U_{\tau i}|$ and $U_{\tau 1} U_{\tau 2}^*\simeq U_{\mu
1}U_{\mu 2}^*$. As a result, at this limit, $P(\nu_\alpha \to
\nu_\tau)$ does not carry new information relative to
$P(\nu_\alpha \to \nu_\mu)$. Muon-track events dominantly receive
contributions from the CC interactions of $\nu_\mu$ and
$\bar{\nu}_\mu$. A subdominant contribution also comes from the CC
interaction of $\nu_\tau$ and the subsequent decay of tau to the
muon which is suppressed by Br$(\tau \to \mu \nu \nu)\simeq 17\%$.
As a result, up to a correction suppressed by Br$(\tau \to \mu \nu
\nu)\sin\theta_{13}$, muon track events within interval
$(t,t+\Delta t)$ from the sharp line is proportional to \be
\label{KT} K(t,\Delta t)\equiv \sum_\alpha n_\alpha\left(\langle
P(\nu_\alpha \to \nu_\mu)\rangle|_{t}^{t+\Delta
t}+\frac{\sigma(\bar{\nu})}{\sigma({\nu})}\langle
P(\bar{\nu}_\alpha \to \bar{\nu}_\mu)\rangle|_{t}^{t+\Delta
t}\right),\ee which should be added to the contribution from the
continuous part of the spectrum. In general, the $\mu$-track
events can be written as \be
\label{ABK}{\mathcal{A}}+{\mathcal{B}}K(t,\Delta t), \ee where
${\mathcal{A}}$ comes from the continuous part of the spectrum. A
priori, both ${\mathcal{A}}$ and ${\mathcal{B}}$ are unknown.
While
$$\phi\equiv \Delta m_{12}^2 L/(2 m_{DM}) \gg 2\pi\ ,$$ the
variation in $\phi$ due to seasonal change in the Sun Earth
distance $\Delta L \sim 5 \times 10^6$ km is of the order of $2
\pi$. This means by studying the seasonal changes in muon track
events, the ratio $\Delta m_{12}^2 / m_{DM}$ can be derived.
\begin{figure}[ht!]
  \centering
  \centerline{\includegraphics[bb=0 0 380 220,keepaspectratio=true,clip=true,angle=0,scale=0.55]{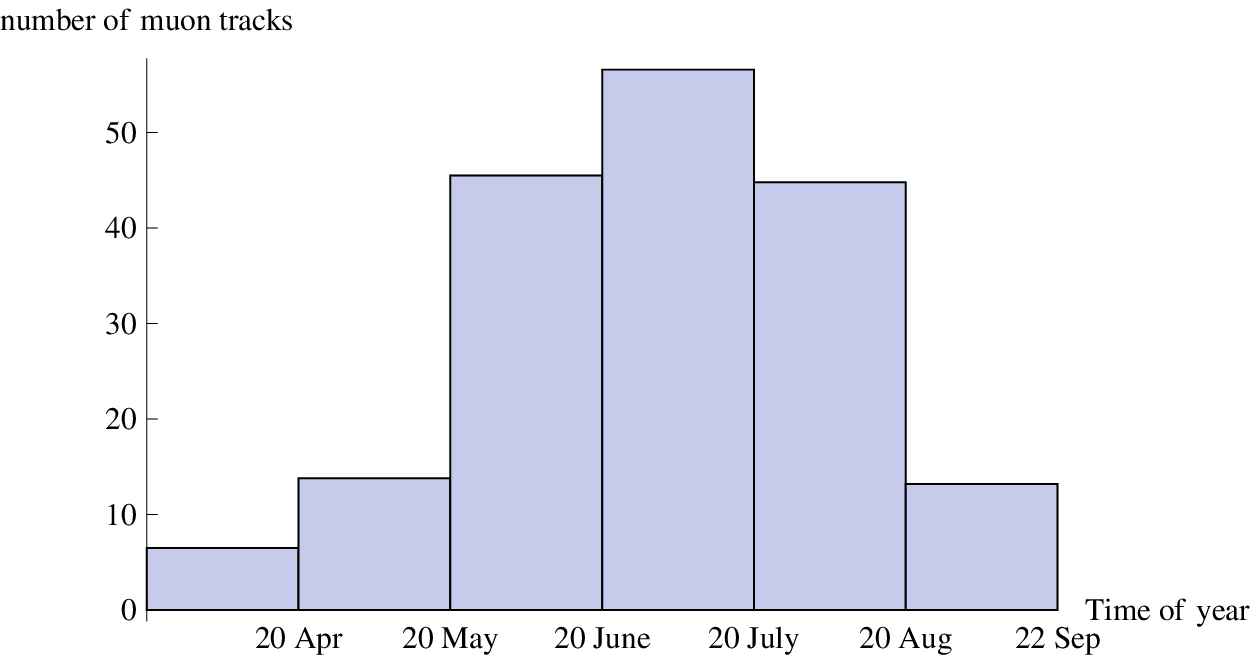}
  \includegraphics[bb=0 0 380
  220,keepaspectratio=true,clip=true,angle=0,scale=0.55]{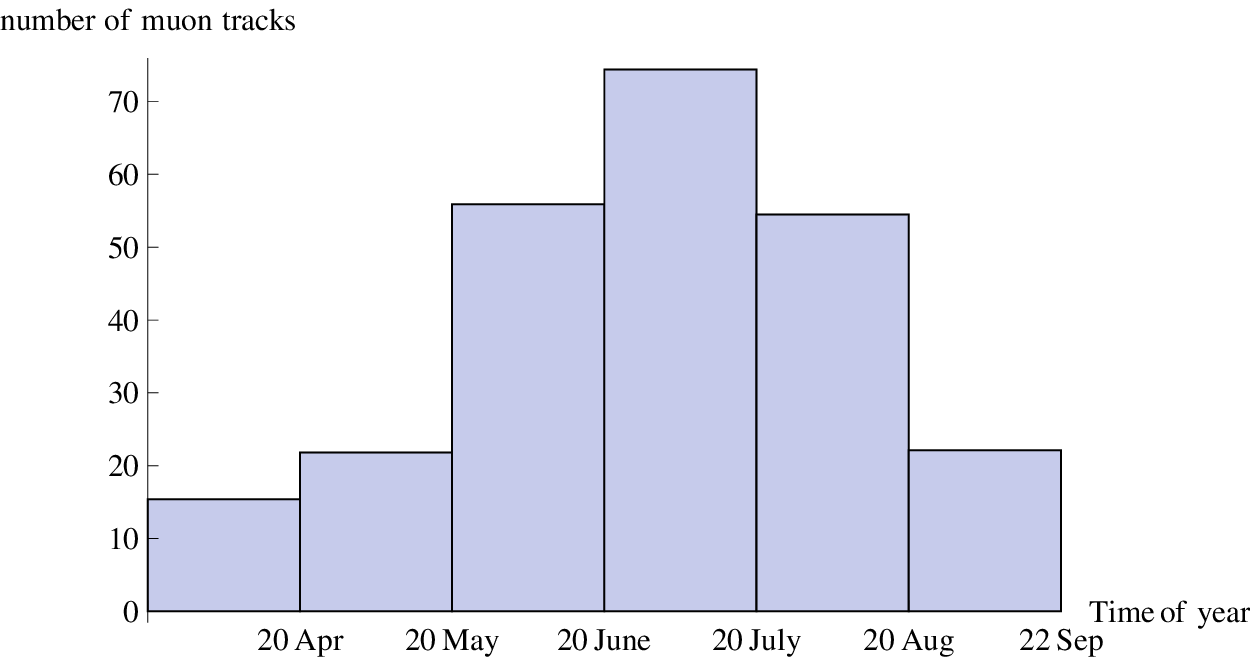}}
 \caption{{\small Number of the through-going $\mu$-track events from
  the spring equinox (20 March) to autumn equinox (22 September) for $\theta_{13}=0$ (left) and
  $\theta_{13}=7^\circ$ (right). In drawing these
  histograms, we have assumed that DM particles with mass $m_{DM}=270$ GeV
  decay via DM+DM$\to\nu_e\bar{\nu}_e$. For the DM capture rate in the Sun,
  we have taken
  $C_\odot=3.4\times 10^{22}$~s$^{-1}$.
 } }
  \label{munumber}
\end{figure}

\begin{figure}[ht!]
  \centering
  \centerline{\includegraphics[bb=0 0 380 220,keepaspectratio=true,clip=true,angle=0,scale=0.55]{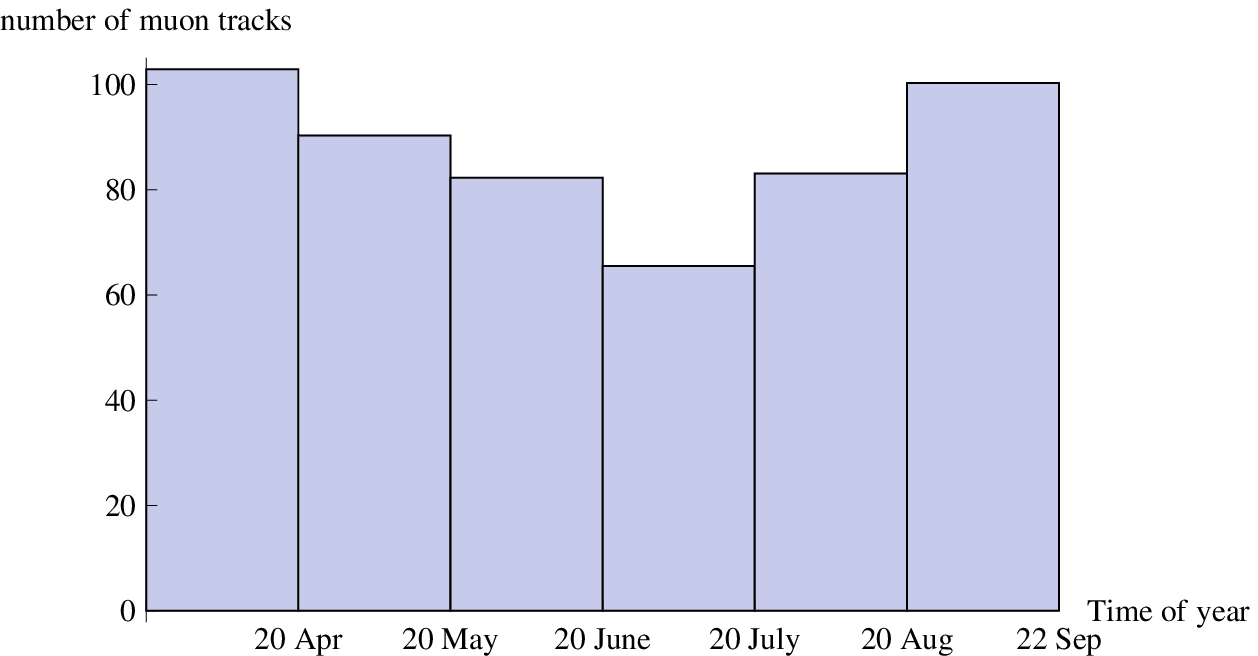}
  \includegraphics[bb=0 0 380
  220,keepaspectratio=true,clip=true,angle=0,scale=0.55]{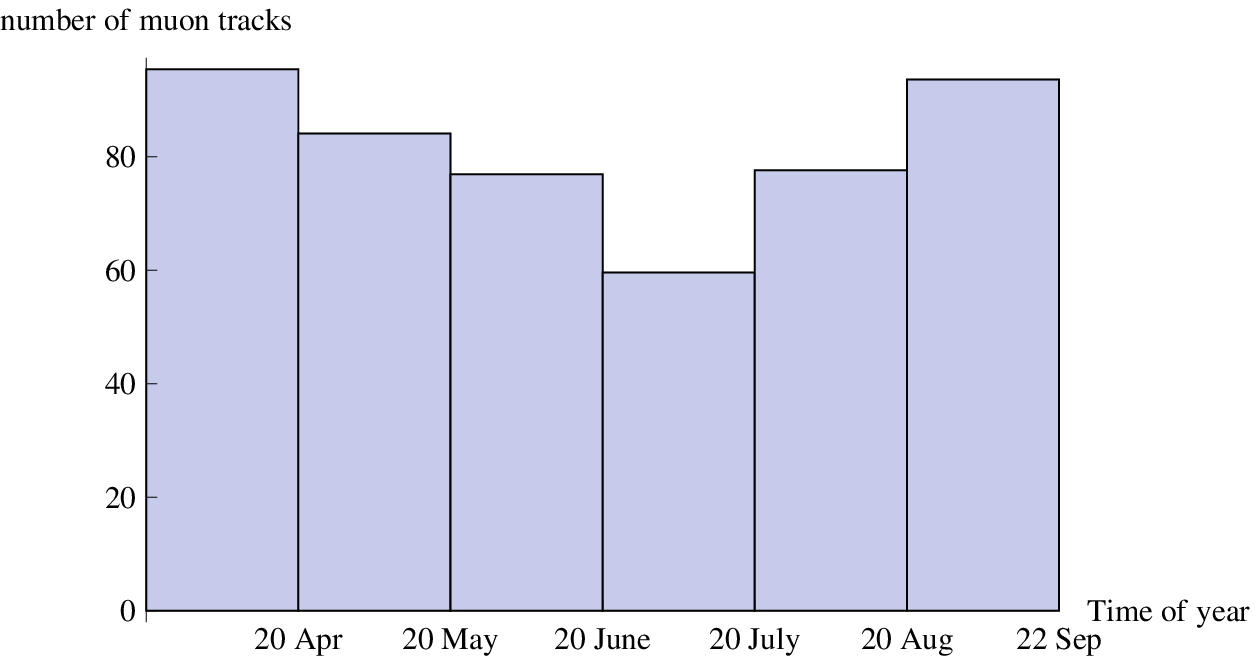}}
 \caption{{\small The same as figure~\ref{munumber} except that DM+DM$\to\nu_\mu\bar{\nu}_\mu$
 } }
  \label{munumbermu}
\end{figure}

Histograms in figures~\ref{munumber} and \ref{munumbermu} display
such seasonal changes for the case that DM pair dominantly
annihilates into neutrino pairs. The horizontal axis shows the
months between spring and autumn equinoxes during which neutrinos
pass the Earth to reach the IceCube. As discussed in the previous
section, during these months atmospheric muon background due to
absorption by Earth is reduced and the data is therefore reliable.
{The peaks and dips in the histograms of figures~\ref{munumber}
and \ref{munumbermu} respectively correspond to the peaks and dips
of the oscillation probabilities $P(\nu_e\to\nu_\mu)$ and
$P(\nu_\mu\to\nu_\mu)$.} If the present bound on the neutrino flux
is saturated, after ten years of collecting data, the number of
events during each month can amount to a few hundred so extracting
$\Delta m_{12}^2/m_{DM}$ with a reasonable accuracy (about 10~\%)
would be possible. Since $\Delta m_{12}^2$ is already known, this
ratio yields the DM mass. The value of the DM mass can in
principle be derived by accelerators such as the LHC. If the two
values coincide, it will be a noteworthy confirmation of the
validity of the approach. In addition to $\Delta m^2 /m_{DM}$, the
seasonal variation also yields \be \label{cte}
{\mathcal{A}}+{\mathcal{B}}\sum_\alpha n_\alpha (|a_{\alpha
i}|^2|U_{\mu i}|^2
+\frac{\sigma(\bar{\nu})}{\sigma(\nu)}|\bar{a}_{\alpha
i}|^2|U_{\mu i}|^2),\ee as well as \be \label{amp}
{\mathcal{B}}\left|\sum_\alpha n_\alpha \left(a_{\alpha 1}^*
a_{\alpha 2} U_{\mu 1}U_{\mu
2}^*+\frac{\sigma(\bar{\nu})}{\sigma(\nu)}\bar{a}_{\alpha 1}^*
\bar{a}_{\alpha 2}U_{\mu 1}^*U_{\mu 2}\right)\right|\ . \ee
Moreover, the seasonal variation yields the following combination
of $\phi$, $\arg[a_{\alpha 1}^* a_{\alpha 2}]$ and
 $\arg[\bar{a}_{\alpha 1}^* \bar{a}_{\alpha 2}]$:
\be \label{phase} \arg\left[\sum_\alpha n_\alpha \left(a_{\alpha
1}^* a_{\alpha 2} U_{\mu 1}U_{\mu
2}^*+\frac{\sigma(\bar{\nu})}{\sigma(\nu)}\bar{a}_{\alpha 1}^*
\bar{a}_{\alpha 2}U_{\mu 1}^*U_{\mu 2}\right)\right]+\phi .\ee
Since $\phi /(2 \pi)\sim 25 (200~{\rm GeV}/m_{DM})$, to derive
meaningful information on $\arg[a_{\alpha 1}^* a_{\alpha 2}]$ and
$\arg[\bar{a}_{\alpha 1}^* \bar{a}_{\alpha 2}]$, $\phi$ has to be
known with percent level accuracy or better which does not seem to
be achievable in foreseeable future. To derive these four
independent pieces of information (i.e., $\Delta
m_{12}^2/m_{DM}$ and three combinations in
eqs.~(\ref{cte},\ref{amp},\ref{phase})), one can divide the time
period between two equinoxes to four segments and measure the
number of $\mu$-tracks in each interval. If the number of events
at each interval exceeds a few 100, the statics will be enough to
perform the analysis and derive the aforementioned combinations
with about 10~\% accuracy.


Notice that the density matrix in eq.~(\ref{nNu}) (or equivalently
matrix $\bar{\rho}$ in eq.~(\ref{nAntinu})) contains seven free
parameters: three real values of $n_\alpha$ and four parameters (3
angles $+$ 1 phase) determining $\nu_\alpha^\prime$ in the flavor
basis. Of course, the three combinations that can be extracted are
not enough to reconstruct $\rho$ (and therefore
${\mathcal{M}}_{\alpha \beta}$). However, these three combinations
yield valuable insight on the flavor structure of $\rho_{\alpha
\beta}$ and therefore ${\mathcal{M}}_{\alpha \beta}$. In the
following, we discuss the additional information that can be
derived from the shower-like events.

As is well-known, shower-like events receive contribution from
three sources: (1) universal NC interaction of all neutrino
species which is insensitive to neutrino oscillation probability
because $\sum_\beta P(\nu_\alpha \to \nu_\beta)=\sum_\beta
P(\bar{\nu}_\alpha \to \bar{\nu}_\beta)=1$; (2) CC interactions of
$\nu_e$ and $\bar{\nu}_e$ whose contributions are given by
$P(\nu_\alpha \to \nu_e)+P(\bar{\nu}_\alpha \to
\bar{\nu}_e)\sigma(\bar{\nu}_e)/\sigma({\nu}_e)$; (3) CC
interactions of $\nu_\tau$ and $\bar{\nu}_\tau$ and the subsequent
hadronic decay of $\tau$ whose effect is given by $[1-{\rm
Br}(\tau \to \mu \nu \nu)][P(\nu_\alpha \to
\nu_\tau)+P(\bar{\nu}_\alpha \to
\bar{\nu}_\tau)\sigma(\bar{\nu}_\tau)/\sigma({\nu}_\tau)]$.
Remember that up to a correction of the order of Max[$\theta_{13},
\theta_{23}-\pi/4$],
 $P(\stackrel{(-)}{\nu_\alpha}\to\stackrel{(-)}{\nu_\mu})
\simeq P(\stackrel{(-)}{\nu_\alpha}\to\stackrel{(-)}{\nu_\tau})$.
One can write the shower-like events as \be
{\mathcal{C}}-r{\mathcal{B}}K(t,\Delta t) \ ,\ee where $K(t ,
\Delta t)$ is defined in eq.~(\ref{KT}) and $r$ is a calculable
quantity (see section \ref{icecube}). The quantity ${\mathcal{C}}$
receives contribution both from continuous part of the spectrum
and the sharp line. Combining the number of shower-like events
with the information derived from the seasonal variation of the
$\mu$-track events yields ${\mathcal{C}}+r{\mathcal{A}}$. In
principle, ${\mathcal{C}}$ can be predicted from ${\mathcal{A}}$
and ${\mathcal{B}}$ but for such a prediction, knowledge of the
shape of the spectrum is required which to some extent can be
extracted from energy measurements. In fact, for a given value of
$m_{DM}$, the shape of continuous spectrum from various
annihilation modes such as ${\rm DM}+{\rm DM} \to b \bar{b}, \tau
\bar{\tau},ZZ, W^+W^-$ can be determined. The shape of spectrum
for scattered and regenerated neutrinos from ${\rm DM}+{\rm DM}
\to \nu\nu$ can be also determined. In other words,
${\mathcal{C}}$ depends on the relative ratios of the different
annihilation modes. Such an analysis might provide a cross-check
for the consistency of the analysis and the measurements. For
further study of the role of $R$ in determining the decay modes
see ref.~\cite{rouzbeh}.

\section{Illustrative examples \label{results}}

From the theoretical point of view, the neutrino states produced
inside the Sun do not need to coincide with any of flavor
eigenstates as the flavor structure of the  physics governing the
DM sector might be different from the physics determining the
flavor structure of the SM. However, for illustrative purposes in
this section, we study the case that the neutrino state at
production corresponds to
$\nu_e$, $\nu_\mu$ or $\nu_\tau$. 
In other words, we assume $\rho$ and $\bar{\rho}$ at production to
be flavor diagonal. Remember that the study applies both
to  ${\rm DM}+{\rm DM} \to \nu_\alpha +\nu_\alpha,
\bar{\nu}_\alpha +\bar{\nu}_\alpha$ and ${\rm DM}+{\rm DM} \to
\nu_\alpha +\bar{\nu}_\alpha$.

At the production point, the matter effect $\sqrt{2} G_F n_e$ is
much larger than $\Delta m_{12}^2/(2 E)$ so $|\nu_e\rangle$ and
$|\bar{\nu}_e\rangle$ will respectively correspond to
$|\nu_2\rangle$ and $|\bar{\nu}_1 \rangle$. As $|\nu_e\rangle$
propagates outside, it will reach the 12-resonance region. At
these energies, the transition is non-adiabatic so at the surface
$|a_{e 1}|\sim |a_{e2}|$. For the case of antineutrinos, there is
no such resonance region so at the surface $|\bar{a}_{e 1}|\gg
|\bar{a}_{e2}|$. For normal hierarchical scheme and nonzero
$\theta_{13}$, neutrinos will undergo a second resonance. For
$\theta_{13}$ within the reach of the forthcoming experiments
\cite{theta13} (i.e., $\theta_{13}>7^\circ$), this resonance
is also non-adiabatic. For inverted hierarchical scheme, instead
of neutrinos, antineutrinos will undergo the resonance due to the
13-splitting.

For lower values of the DM mass, detection rate becomes smaller as
the neutrino-nuclei cross section decreases with lowering the
energies of neutrinos. On the other hand, with increasing
$m_{DM}$, the sharp line composed of the unscattered neutrinos
diminishes, leading to a suppressed $\Delta$.
It can be shown that  for $m_{DM}\gtrsim500$~GeV, the tail of the
spectrum from the scattered neutrinos  suppresses $\Delta$ down to
values $\Delta\lesssim 0.5$. Thus, generally we expect that in the
range, $100~{\rm GeV}<m_{DM}<500~{\rm GeV}$,  the $\Delta$
measurement method will be most useful. We focus on $m_D\simeq
200$ GeV and study the high sensitivity of $\Delta$ to $m_{DM}$.
We have checked for other values of $m_{DM}$ in the range  100~GeV
to 300~GeV and found that our results are robust against varying
$m_{DM}$. Using the code described in section~\ref{num}, we have
calculated the seasonal variation, $R_{{\rm DC}}$ and $R_{{\rm
thr}}$ for $m_{DM}=197,200,203$~GeV, $\theta_{13}=0,7^\circ$ and
$\delta=0,\pi/2,\pi$, taking the initial neutrino flavor to be
$\nu_e$, $\nu_\mu$ and $\nu_\tau$. The results are displayed in
tables \ref{nuetable}, \ref{numutable} and \ref{nutautable}.

For the shower-like events, the effective volume of a detector is
approximately equal to its geometrical volume: $V_{eff}^{{\rm
shower}}=d A_{eff}$ where $d$ and $A_{eff}$ are respectively the
depth and effective area of the detector. However, since the muons
can penetrate farther, the effective volume for their detection
can be larger: $V_{eff}^{\mu}=(R_\mu+d) A_{eff}$ where $R_\mu$ is
the muon range which depends only on the chemical composition of
the medium. As a result, the ratio of $\mu$-track to shower-like
events is enhanced by a factor of $(R_\mu +d)/d$. Obviously, for a
smaller detector this enhancement factor is larger. That is why
the ratio in the case of Deepcore, $R_{{\rm DC}}$, is so much
larger than the ratio in the case of the whole detector, $R_{{\rm
thr}}$. (See tables \ref{nuetable}, \ref{numutable} and
\ref{nutautable}.)

\begin{table}
\begin{center}
\begin{tabular}{|c|c|c|c|||c|c|c|c|c|c|}
\hline $m_{DM}$& N/I & $\theta_{13}$& $\delta$ & $R^{{\rm IC}}$ &
$R^{{\rm{DC}}}$ & $\Delta^{{\rm{IC}}}_{{\rm{20Mar}}}$&
$\Delta^{{\rm{IC}}}_{{\rm{3Apr}}}$&
$\Delta^{{\rm{DC}}}_{{\rm{20Mar}}}$& $\Delta^{{\rm{DC}}}_{{\rm{3Apr}}}$\\
\hline
 197 & N & 0 & 0 & 0.4 & 14 & 0.7& 0.7 & 0.7 & 0.7 \\
197 & I & 0 & 0 & 0.4 & 14 & 0.7 & 0.7 & 0.7 & 0.7 \\
197 & N & 7$^\circ$ & 0 & 0.5 & 18 & 0.5 & 0.6 & 0.6 & 0.6 \\
197 & I & 7$^\circ$ & 0 & 0.6 & 20& 0.6 & 0.6 & 0.6 & 0.6 \\
197 & N & 7$^\circ$ & $\pi/2$ & 0.5 & 18 & 0.5 & 0.5 & 0.5 & 0.5 \\
197 & I & 7$^\circ$ & $\pi/2$ & 0.6 & 19 & 0.5 & 0.5 & 0.5 & 0.5 \\
197 & N & 7$^\circ$ & $\pi$ &  0.4 & 14 & 0.6 & 0.6 & 0.6 & 0.6 \\
 197 & I & 7$^\circ$ & $\pi$ & 0.5 & 16 & 0.4 & 0.5 & 0.4 & 0.5 \\ \hline
200& N & 0 & 0 & 0.5 & 15 & 0.5 & 0.5 & 0.5 & 0.5 \\
 200 & I & 0 & 0 & 0.5 & 15 & 0.5 & 0.5 & 0.5 & 0.5 \\
 200 & N & 7$^\circ$ & 0 & 0.6 & 19 & 0.4 & 0.5 & 0.4 & 0.4 \\
 200 & I & 7$^\circ$ & 0 & 0.7 & 22 & 0.4 & 0.4 & 0.3 & 0.4 \\
 200& N & 7$^\circ$ & $\pi/2$ & 0.6& 17& 0.4 & 0.5 & 0.4 & 0.4 \\
200& I & 7$^\circ$ & $\pi/2$ & 0.7 & 19& 0.4 & 0.4 & 0.4 & 0.4 \\
200& N & 7$^\circ$ & $\pi$ & 0.5 & 14 & 0.4&0.4 & 0.4 & 0.4 \\
200& I & 7$^\circ$ & $\pi$ & 0.5 & 15 & 0.4 & 0.4 & 0.4 & 0.4 \\
\hline
203 & N & 0 & 0 & 0.5  & 15 & 0.1 & 0.2 & 0.2 & 0.2 \\
203 & I & 0 & 0 & 0.5 & 15 & 0.1 & 0.2 & 0.1 & 0.2 \\
203 & N & 7$^\circ$ & 0 & 0.6 & 19 & 0.1 & 0.1 & 0.2 & 0.2 \\
203 & I & 7$^\circ$ & 0 & 0.7 & 20 & 0.1 & 0.1 & 0.1 & 0.1 \\
203 & N & 7$^\circ$ & $\pi/2$ & 0.6 & 19 & 0.0 & 0.0 & 0.1 & 0.1 \\
203 & I & 7$^\circ$& $\pi/2$ & 0.6 & 19 & 0.0 & 0.1 & 0.1 & 0.1 \\
203 & N & 7$^\circ$ & $\pi$ &  0.5 & 15  & 0.1 & 0.1 & 0.2 & 0.2 \\
 203 & I & 7$^\circ$ & $\pi$ & 0.6 & 17 & 0.1 & 0.1 & 0.2 & 0.2 \\ \hline
 \hline
\end{tabular}
\caption{Seasonal variation and $\mu$-track to shower-like ratio
for DM particles with mass $m_{DM}$ (in GeV) and annihilation mode ${\rm DM}+{\rm DM} \to \nu_e {\nu}_e$. N/I indicates normal
versus inverted neutrino mass scheme. $R^{{\rm IC}}$ indicates the
ratio of the numbers of muon-track to shower-like event  for
through-going events measured by the whole IceCube. $R^{{\rm DC}}$
is the same quantity measured by DeepCore. $\Delta_{{\rm{20Mar}}}$
is the seasonal variation between two equinoxes
$\Delta_{{\rm{20Mar}}}\equiv \Delta(20~{\rm Mar},186~{\rm
days},23~{\rm Sep},179~{\rm days})$. Finally,
$\Delta_{{\rm{3Apr}}}\equiv \Delta(3~{\rm Apr},186~{\rm
days},6~{\rm Oct},179~{\rm days}).$ See eqs.
(\ref{deltaEQUIV},\ref{DCmu},\ref{Rdefinition}).} \label{nuetable}
\end{center}
\end{table}
\begin{table}
\begin{center}
\begin{tabular}{|c|c|c|c|||c|c|c|c|c|c|}
\hline $m_{DM}$& N/I & $\theta_{13}$& $\delta$ & $R^{{\rm IC}}$ &
$R^{{\rm DC}}$ & $\Delta^{{\rm{IC}}}_{{\rm{20Mar}}}$& $\Delta^{{\rm{IC}}}_{{\rm{3Apr}}}$&
$\Delta^{{\rm{DC}}}_{{\rm{20Mar}}}$& $\Delta^{{\rm{DC}}}_{{\rm{3Apr}}}$\\
\hline
197 & N & 0 & 0 & 1.0 & 47 & 0.1 & 0.1 & 0.0 & 0.1 \\
197 & I & 0 & 0 & 1.0 & 48 & 0.1 & 0.1 & 0.0 & 0.1 \\
197 & N & 7$^\circ$ & 0 & 0.9 & 42 & 0.1 & 0.1 & 0.1 & 0.1 \\
197 & I & 7$^\circ$ & 0 & 0.9 & 42& 0.1 & 0.1 & 0.1 & 0.1 \\
197 & N & 7$^\circ$ & $\pi/2$ & 0.9 & 44 & 0.1 & 0.1 & 0.1 & 0.1 \\
197 & I & 7$^\circ$ & $\pi/2$ & 1.0& 48 & 0.0 & 0.0 & 0.0 & 0.0 \\
197 & N & 7$^\circ$ & $\pi$ & 1.0 & 51 & 0.0 & 0.0 & 0.0 & 0.0 \\
 197 & I & 7$^\circ$ & $\pi$ & 1.0 & 50 & 0.0 & 0.0 & 0.0 & 0.0 \\\hline
200 & N & 0 & 0 & 1.0 & 49 & 0.1 & 0.1 & 0.1 & 0.1 \\
200 & I & 0 & 0 & 1.0 & 47 & 0.1 & 0.1 & 0.0 & 0.0 \\
200 & N & 7$^\circ$ & 0 & 0.9 & 43 & 0.1 & 0.1 & 0.05 & 0.05 \\
200 & I & 7$^\circ$ & 0 & 0.8 & 41 & 0.1 & 0.2 & 0.1 & 0.1 \\
200 & N & 7$^\circ$ & $\pi/2$ & 0.9 & 46 & 0.1 & 0.1 & 0.1 & 0.1 \\
200 & I & 7$^\circ$ & $\pi/2$ & 1.0 & 48 & 0.1& 0.1 & 0.0 & 0.0 \\
200 & N & 7$^\circ$ & $\pi$ & 1.0 & 52 & 0.1 & 0.1 & 0.1 & 0.1 \\
 200 & I & 7$^\circ$ & $\pi$ & 1.0 & 50 & 0.0 & 0.0 & 0.0 & 0.0 \\
\hline
203 & N & 0 & 0 & 1.0 & 50 & 0.0 & 0.0 & 0.0 & 0.0 \\
203 & I & 0 & 0 & 1.1  & 52 & 0.0 & 0.0 & 0.0 & 0.0 \\
203 & N & 7$^\circ$ & 0 & 0.9 & 45 & 0.1 & 0.1 & 0.0 & 0.1 \\
203 & I & 7$^\circ$ & 0 & 1.0 & 46 & 0.1 & 0.1 & 0.1 & 0.1 \\
203 & N & 7$^\circ$ & $\pi/2$ & 1.0 & 47 & 0.0 & 0.0 & 0.0 & 0.0 \\
203 & I & 7$^\circ$ & $\pi/2$ & 1.1 & 52 & 0.0 & 0.0 & 0.0 & 0.0 \\
203 & N & 7$^\circ$ & $\pi$ & 1.1 & 54 & 0.0 & 0.0 & 0.0 & 0.0 \\
 203 & I & 7$^\circ$ & $\pi$ &1.0 & 52 & 0.1 & 0.1 & 0.0 & 0.0 \\ \hline
 \hline
\end{tabular}
 \caption{The same as Table \ref{nuetable} for
DM+DM$\to \nu_\mu \nu_\mu$.}\label{numutable}
\end{center}
\end{table}
\begin{table}
\begin{center}
\begin{tabular}{|c|c|c|c|||c|c|c|c|c|c|}
\hline $m_{DM}$& N/I & $\theta_{13}$& $\delta$ & $R^{{\rm IC}}$ &
$R^{{\rm{DC}}}$ & $\Delta^{{\rm{IC}}}_{{\rm{20Mar}}}$& $\Delta^{{\rm{IC}}}_{{\rm{3Apr}}}$&
$\Delta^{{\rm{DC}}}_{{\rm{20Mar}}}$& $\Delta^{{\rm{DC}}}_{{\rm{3Apr}}}$\\
\hline
197 & N & 0 & 0 & 1.1 & 65 & 0.1 & 0.1 & 0.1 & 0.1 \\
197 & I & 0 & 0 & 1.1 & 65 & 0.1 & 0.1 & 0.1 & 0.1 \\
197 & N & 7$^\circ$ & 0 & 1.0 & 59 & 0.1 & 0.1 & 0.1 & 0.1 \\
197 & I & 7$^\circ$ & 0 & 1.0 & 57 & 0.1 & 0.1 & 0.1 & 0.1 \\
197 & N & 7$^\circ$ & $\pi/2$ & 1.1 & 64 & 0.1 & 0.1 & 0.1 & 0.1 \\
197 & I & 7$^\circ$ & $\pi/2$ & 1.0 & 59 & 0.1 & 0.1 & 0.1 & 0.1 \\
197 & N & 7$^\circ$ & $\pi$ & 1.1  & 68 & 0.1 & 0.1 & 0.1 & 0.1 \\
 197 & I & 7$^\circ$ & $\pi$ & 1.1 & 66 & 0.1 & 0.1 & 0.0 & 0.1 \\
\hline
200 & N & 0 & 0 & 1.1 & 66 & 0.1 & 0.1 & 0.0 & 0.0 \\
200 & I & 0 & 0 & 1.1 & 67 & 0.1 & 0.1 & 0.0 & 0.0 \\
200 & N & 7$^\circ$ & 0 & 1.0 & 59 & 0.1 & 0.1 & 0.0 & 0.0 \\
200 & I & 7$^\circ$ & 0 & 1.0 & 58 & 0.0 & 0.0 & 0.0 & 0.0 \\
200 & N & 7$^\circ$ & $\pi/2$ & 1.1 & 65 & 0.0 & 0.0 & 0.0 & 0.0 \\
200 & I & 7$^\circ$ & $\pi/2$ & 1.0 & 61 & 0.1 & 0.1 & 0.0 & 0.0 \\
200 & N & 7$^\circ$ & $\pi$ & 1.1 & 70 & 0.0 & 0.0 & 0.0 & 0.0 \\
 200 & I & 7$^\circ$ & $\pi$ & 1.1 & 69 & 0.1 & 0.1 & 0.0 & 0.0 \\
\hline
203 & N & 0 & 0 & 1.1 & 67 & 0.0 & 0.0 & 0.0 & 0.0 \\
203 & I & 0 & 0 & 1.0 & 66 & 0.0 & 0.0 & 0.0 & 0.0 \\
203 & N & 7$^\circ$ & 0 & 1.0 & 60 & 0.0 & 0.0 & 0.0 & 0.0 \\
203 & I & 7$^\circ$ & 0 & 0.9 & 58 & 0.1 & 0.1 & 0.0 & 0.0 \\
203 & N & 7$^\circ$ & $\pi/2$ & 1.0 & 66 & 0.0 & 0.0 & 0.0 & 0.0 \\
203 & I & 7$^\circ$ & $\pi/2$ & 0.9 & 61 & 0.0 & 0.0 & 0.0 & 0.0 \\
203 & N & 7$^\circ$ & $\pi$ & 1.1 & 71 & 0.0 & 0.0 & 0.0 & 0.0 \\
 203 & I & 7$^\circ$ & $\pi$ & 1.0 & 69 & 0.0 & 0.0 & 0.0 & 0.0 \\ \hline
 \hline
\end{tabular}
\caption{The same as Table \ref{nuetable} for DM+DM$\to \nu_\tau
\nu_\tau$.}\label{nutautable}
\end{center}
\end{table}

The matter effects inside
 the Sun do not distinguish between $\nu_\mu$ and $\nu_\tau$. Thus,
 if the neutrino mass matrix is $\mu\tau$-symmetric in vacuum
 (i.e., $\theta_{13}=\theta_{23}-\pi/4=0$), the neutrino sector remains $\mu\tau$-symmetric in the matter. This means that in this limit, regardless of whether
neutrino mass scheme is normal or inverted, we can write \be
\label{NUE} |\nu_e;{\rm surface}\rangle=\cos
\theta_{12}^s|1\rangle + \sin \theta_{12}^s|2\rangle \ ,\ee \be
|\nu_\mu;{\rm surface}\rangle=-\frac{\sin
\theta_{12}^s}{\sqrt{2}}|1\rangle + \frac{\cos
\theta_{12}^s}{\sqrt{2}}|2\rangle +\frac{1}{\sqrt{2}}|3\rangle \ee
and
 \be
|\nu_\tau;{\rm surface}\rangle=\frac{\sin
\theta_{12}^s}{\sqrt{2}}|1\rangle - \frac{\cos
\theta_{12}^s}{\sqrt{2}}|2\rangle +\frac{1}{\sqrt{2}}|3\rangle \
.\ee Notice that $\theta_{12}^s$ is different from $\theta_{12}$
mixing angle in the  $U_{{\rm MNS}}$ neutrino mixing matrix. Similar relations
holds for antineutrinos but with a different value of
$\bar{\theta}_{12}^s$. For antineutrinos, we expect  $\cos
\bar{\theta}_{12}^s\simeq 1$. In terms of the notation in eqs.
(\ref{timed},\ref{antiExpansion}), $a_{e1}=\cos \theta_{12}^s$,
$\bar{a}_{e1}=\cos \bar{\theta}_{12}^s$, $a_{e3}=\bar{a}_{e3}=0$,
$a_{\mu 3}=\bar{a}_{\mu 3}=a_{\tau 3}=\bar{a}_{\tau 3}=1/\sqrt{2}$
and etc. From the above relations we find
 $|\stackrel{(-)}{a_{\mu i}}|=|\stackrel{(-)}{a_{\tau i}}|$,
 $a_{\mu 1}^* a_{\mu 2} U_{\mu 1}U_{\mu 2}^*
=a_{\tau1}^* a_{\tau2} U_{\mu 1}U_{\mu 2}^*$ and  $\bar{a}_{\mu
1}^* \bar{a}_{\mu 2} U_{\mu 1}^*U_{\mu 2} =\bar{a}_{\tau1}^*
\bar{a}_{\tau2} U_{\mu 1}^*U_{\mu 2}$. Notice however that $a_{\mu
1}^* a_{\mu 2} U_{\mu 1}U_{\mu 2}^*\ne a_{\tau1}^* a_{\tau3}
U_{\mu 1}U_{\mu 3}^*$. Thus, in the $\mu \tau$-symmetric case, up
to the corrections of order of $O_{13}$ (see eq.~(\ref{ooo13}) for
the definition), $\langle P(\nu_\mu \to \nu_{\mu})\rangle\simeq
\langle P(\nu_\tau \to \nu_\mu)\rangle$ and $\langle
P(\bar{\nu}_\mu \to \bar{\nu}_{\mu})\rangle\simeq \langle
P(\bar{\nu}_\tau \to \bar{\nu}_\mu)\rangle$. This means, to
leading approximation, if we take only the un-scattered neutrinos,
the contribution to $\Delta$ and $R$ will be the same for initial
$ \nu_\mu$ and $\nu_\tau$. This makes distinguishing between
$({\rm DM}+{\rm DM}\to \nu_\mu \nu_\mu)$ and $({\rm DM}+{\rm DM}
\to \nu_\tau \nu_\tau)$ difficult. However, the regenerated
neutrinos coming from $\nu_\tau \to \tau \to \nu_\tau$ in the Sun
differentiate between them. Remember that at these energies
$\Delta m_{13}^2 R_{\odot}/2E<2 \pi$. In fact, $\nu_\mu$ can only
partially oscillate to $\nu_\tau$ before exiting the Sun. Thus,
the contribution to regenerated neutrinos strongly depends on the
initial flavor of  the neutrino. Of course the regenerated
neutrinos will have less energy, so they are more important for
DeepCore than the whole IceCube because the energy threshold of
DeepCore is lower. Our results show that for
$\theta_{13}=\theta_{23}-\pi/4=0$, the deviations of $\Delta^{{\rm
IC}}$ and $R^{{\rm IC}}$  for initial $\nu_\mu$ from the same
quantities for initial $\nu_\tau$ are negligible (compare tables
\ref{numutable} and \ref{nutautable}). As seen from the tables,
this deviation for the case measured by the DeepCore (i.e.,
$\Delta^{{\rm DC}}$ and $R^{{\rm IC}}$) is larger. This can be
explained by the contribution from the regenerated neutrinos.

Another point to note is that since the validity of
eq.~(\ref{NUE}) does not depend on scheme, we expect that for
$\theta_{13}=0$ and initial $\nu_e$, $\Delta$ and $R$ to be the
same for normal and inverted mass schemes. Results shown in
table~\ref{nuetable} confirm this expectation.

The tables show that the sensitivity of $\Delta$ to $m_{DM}$ is
very high. As discussed in the previous section and shown in
figure~\ref{munumber}, by studying the variation over four intervals
of time, $\Delta m_{12}^2/m_{DM}$ can be derived.

From the tables we observe that for $({\rm DM}+{\rm DM}\to
\nu_e+\nu_e)$, $\Delta$ can be significant; however, for the cases
$({\rm DM}+{\rm DM}\to \nu_\mu+\nu_\mu)$ and $({\rm DM}+{\rm
DM}\to \nu_\tau+\nu_\tau)$, $\Delta$ is in general small. We found
that this observation is robust against varying $m_{DM}$. The
reason is two folded: (i) For initial $\nu_\mu$ and $\nu_\tau$,
the contribution from the continuous regenerated neutrinos to the
flux at the detector is higher so the denominator of the ratio in
eq. (\ref{deltaEQUIV}) is enhanced leading to lower values of
$\Delta$. From tables \ref{numutable} and \ref{nutautable}, we
observe that $\Delta^{{\rm DC}}$ is smaller than $\Delta^{{\rm
IC}}$. The reason is that DeepCore has a lower detection threshold
so it receives a larger contribution from regenerated neutrinos
which are in general less energetic. (ii) The second reason is
that for $\alpha=\mu, \tau$; $a_{\alpha 1} a_{\alpha 2} U_{\mu 1}
U_{\mu 2}$ is smaller (see eq. (\ref{average})). However, we
should remember that $({\rm DM}+{\rm DM}\to \nu_\mu+\nu_\mu)$ and
$({\rm DM}+{\rm DM}\to \nu_\tau+\nu_\tau)$ are quite specific
cases. In general DM pair can annihilate to  general coherent
combinations of the different neutrino flavors for which $\Delta$
is in general large.

\section{Conclusion and discussion\label{conclusion}}

We have shown that by studying the time variation of $\mu$-track
events, valuable information on the nature of DM particles can be
derived: (i) Measuring a nonzero value for $\Delta$ defined in
eq.~(\ref{deltaEQUIV}) implies that there is a sharp monochromatic
feature in the neutrino spectrum  which in turn means ${\rm DM+DM}
\to \nu +\nu$ is [one of] the significant annihilation modes. (ii)
Once such a feature is established, by studying the $\mu$-track
events over four different time intervals, the value of $\Delta
m_{12}^2/m_{DM}$ as well as three independent combinations  (see
eqs. (\ref{cte},\ref{amp},\ref{phase}) as well as eq. (\ref{ABK}))
of $a_{\alpha i}$ and $\bar{a}_{\alpha j}$ which are defined in
eqs. (\ref{timed},\ref{antiExpansion}) can be extracted. Thus, by
this method, $m_{DM}$ can be extracted and checked against the
value derived from studying the endpoint of the spectrum or from
the accelerator experiments. Although the information will not be
enough to completely reconstruct the flavor structure of the
amplitude of ${\rm DM+DM}\to \nu_\alpha +\nu_\beta$, the
combinations in eqs.~(\ref{cte}) and (\ref{amp}) contain
invaluable information that can constrain  models predicting ${\rm
DM+DM}\to \nu_\alpha +\nu_\beta$. In particular, observing
nonzero oscillatory effects rules out models predicting democratic
neutrino flavor production. Moreover, our numerical results show
that by studying $\Delta$, models predicting ${\rm DM}+{\rm DM}
\to \nu_e+ \stackrel{(-)}{\nu_e}$ can be distinguished from those
predicting  ${\rm DM}+{\rm DM} \to \nu_\mu+
\stackrel{(-)}{\nu_\mu}$ or
 ${\rm DM}+{\rm DM}
\to \nu_\tau +\stackrel{(-)}{\nu_\tau}$. Notice that to derive
information on $\mathcal{M}_{\alpha \beta}$  from $a_{\alpha i}$,
the evolution of the neutrino state from the Sun center to its
surface has to be taken into account. This can be readily done by
codes such as the one developed to derive the numerical results of
the present paper ({\emph{see}}, tables \ref{nuetable},
\ref{numutable}, \ref{nutautable}).

We showed that it is possible that while DM particles dominantly
annihilate into neutrino pairs, their interaction rate with nuclei
in the Sun is high enough to lead to high enough DM capture rate
and therefore to a neutrino flux saturating the present bounds. If
the present bound on the neutrino flux from dark matter
annihilation is saturated, after about 10 years of data taking,
the statistics will be high enough to make this method reliable.
In this method, only $\mu$-track events are employed so high
energy threshold for the detection of shower-like events will not
be a hinderance. To carry out this analysis, it will be enough to
take data from 20th of March to 23rd of September when the
neutrino from the Sun pass through the Earth to reach the detector
and as a result, the atmospheric muon background will not be a
problem. The method presented in this paper can be combined with
the energy spectrum measurements \cite{Strumia} and the ratio of
$\mu$-track to shower-like events \cite{rouzbeh} to draw a more
complete picture of the dark matter annihilation modes.


\acknowledgments We are grateful to R.~Allahverdi and
K.~Richardson-McDaniel for useful discussion through which the
idea of this paper was born as well as for collaborating at the
early stages of this work. We would also like to thank M.~Blennow
and T.~Ohlsson for useful discussions. We specially appreciate
Prof. F.~Halzen for his encouragements. A.~E. appreciates
``Bonyad-e-Melli-e-Nokhbegan'' in Iran and the Brazilian funding
agency Funda\c{c}\~ao de Amparo \`a Pesquisa do Estado de S\~ao
Paulo (FAPESP) for partial financial support.


\end{document}